\documentclass[vecphys]{svmult}

\usepackage{makeidx}         
\usepackage{graphicx}        
\usepackage{multicol}        
\usepackage[bottom]{footmisc}

\newcommand{\aap}{A\&A}

\newcommand{\aj}{AJ}
\newcommand{\apj}{ApJ}

\newcommand{\apjs}{ApJS}

\newcommand{\mnras}{MNRAS}

\def \kms{\ifmmode{~{\rm km\,s}^{-1}}\else{~km~s$^{-1}$}\fi}
\def \vhel{\ifmmode{V_{{\rm hel}}}\else{$V_{{\rm hel}}$}\fi}
\def \vsys{\ifmmode{V_{{\rm sys}}}\else{$V_{{\rm sys}}$}\fi}
\def \vobs{\ifmmode{V_{{\rm obs}}}\else{$V_{{\rm obs}}$}\fi}
\def \degree{\ifmmode{^{\circ}}\else{$^{\circ}$}\fi}
\def \lsun{\ifmmode{{\rm\ L}_\odot}\else{${\rm\ L}_\odot $}\fi}
\def \msun{\ifmmode{{\rm\ M}_\odot}\else{${\rm\ M}_\odot$}\fi}
\def \myr{\ifmmode{{\rm\ M}_\odot{\rm\ yr}^{-1}}\else{${\rm\ M}_\odot$ 
yr$^{-1}$}\fi}
\def \teff{\ifmmode{{\rm{T}}_{\rm eff}}\else{${\rm{T}}_{\rm eff}$}\fi}
\def \mdot{\ifmmode{{\rm\dot{M}}}\else{${\rm\dot{M}}$}\fi}

\newcommand{\ha}{H$\alpha$}
\newcommand{\hb}{H$\beta$}

\newcommand{\heii}{He\,{\sc ii}}
\newcommand{\heiil}{He\,{\sc ii}\ 6560\,\AA}
\newcommand{\heiis}{He\,{\sc ii}\ 4686\,\AA}

\newcommand{\oiii}{[O\,{\sc iii}]}
\newcommand{\oiiil}{[O\,{\sc iii}]\ 5007\,\AA}

\newcommand{\nii}{[N\,{\sc ii}]}
\newcommand{\niil}{[N\,{\sc ii}]\ 6584\,\AA}

\def \st{\ifmmode{^{\mathrm{st}}}\else{${^{\mathrm{st}}}$}\fi}
\def \nd{\ifmmode{^{\mathrm{nd}}}\else{${^{\mathrm{nd}}}$}\fi}
\def \rd{\ifmmode{^{\mathrm{rd}}}\else{${^{\mathrm{rd}}}$}\fi}
\def \th{\ifmmode{^{\mathrm{th}}}\else{${^{\mathrm{th}}}$}\fi}

\makeindex             

\begin{document}

\title*{Do fast winds dominate the dynamics of planetary nebulae?}
\titlerunning{Fast winds in PNe} 
\author{John Meaburn\inst{1}}
\institute{Instituto de Astronomia, UNAM, Apdo. Postal 877, Ensenada, BC 22800,
M{\'exico}.
\texttt{jm@ast.man.ac.uk}}

\maketitle

\section{Introduction}

{\bf A review of recent observations of the kinematics of six objects 
that represent
the broad range of phenomena called planetary nebulae (PNe)
is presented. It is demonstrated that Hubble--type outflows are 
predominant, consequently it 
is argued that ballistic
ejections from the central stars could have dominated the dynamical effects 
of the fast winds in several, and perhaps all, of these objects. 
An alternative possibility, which involves an extension to the
Interacting Winds model,
is considered to explain the dynamics of evolved planetary nebulae.}

A consensus has been established (e.g. Kastner et al. 2003) 
about the basic processes for the creation of a planetary nebula (PN):
an intermediate mass star (initial mass 1--8 \msun) loses mass in 
its Asymptotic Giant Branch (AGB) phase at 
$\leq$ 10$^{-4}$ \msun\ yr$^{-1}$ by emitting a `superwind' 
flowing at 10 - 20 \kms\ over $\leq$ 10$^{5}$ yr (depending
on the initial stellar mass).
The star eventually becomes an 0.5 - 1.0 \msun\ White Dwarf (WD) which produces
enough Lyman photons to ionise a substantial fraction of 
the circumstellar envelope recognisable
as the expanding PN. The whole structure
can be enveloped as well in a prior low density Red Giant (RG), similarly slow,
wind. 
In the transition from the AGB to WD phase the 
outflow mass loss rate declines to 10$^{-8}$ \msun\ yr$^{-1}$ but  
increases its 
speed dramatically to several 1000 \kms\ to blow as a fast wind
for an as yet unknown period. As the WD star evolves further the fast
wind declines. 

Obviously the real story is hugely more complicated in detail and variable
between objects (Balick \& Frank, 2002). Morphologies range from simple, 
spherical shells
to complex poly-polar structures (e.g. NGC 2440, L{\'o}pez et al. 1998) 
probably around close binary systems. The ejected, 
dusty, AGB, molecular superwind 
material is often very clumpy (e.g. the cometary
globules of NGC 7293, Huggins et al
1992; Meaburn et al. 1992; O'Dell \& Handron 1996; Meaburn et al. 1998).  
High--speed jets (e.g. IRAS17423-1755, Riera et al. 1995) and `bullets' 
(e.g. MyCn 18, Bryce et al. 1997; O'Connor et al. 2000) are 
found and even shell or lobe
expansion velocities
can range from  20 \kms\ to $\geq$ 500 \kms\ (e.g. He2--111, Meaburn \& Walsh
1989 and NGC 6302, Meaburn et al. 2005c). Sometimes many of these
distinctly separate phenomena occur in one object
reflecting the separate stages of its complex evolution. The ages of 
observed PNe range from
the initial proto-PN stage to those of well--evolved PN $\geq$~10$^{4}$ yr 
later around an
ageing WD star. 

In current dynamical theories of PNe  much emphasis is placed on
the importance, even dominance, of the fast wind. In the elegant interacting
winds (IWs) model and its variants 
(Kwock, Purton \& Fitzgerald 1978; Kahn \& West 1985;
Chu et al. 1993; Mellema 1995 \& 1997; Balick \& Frank 2002) the shocked 
(10${^6}$ -- 10$^{8}$ K)
fast wind can form an energy--conserving, 
pressure--driven `bubble' in the preceding smooth AGB wind
whose density declines as distance$^{-2}$ from the star.
This possibility is similar in principle to that pioneered 
by Dyson \& de Vries (1972) albeit 
within  a stationary 
medium of uniform density. The characteristic shell of a simple PN 
is consequently formed between the shocks in the fast wind and 
AGB wind
and, being pressure--driven by the superheated gas, 
is expanding faster than this
ambient AGB outflow. 
A variant would have the momentum of the isotropic fast wind
simply sweeping up the AGB outflow and accelerating an expanding shell. 
For the creation of a  bi-polar PN Cant{\'o} (1978) and
Barral \& Cant{\'o} (1981) considered something similar. Here a fast
wind from a star embedded in a dense circumstellar
disk forms cavities  on either side of it which are
delineated by stationary shocks across which the fast wind refracts
to form bi--polar, momentum--conserving, outflows parallel to the cavity walls.
Again, energy conserving, elongated `bubbles', pressure driven by the
shocked wind 
on either side of this disk, would also form expanding bi-polar lobes.
Steffen \& L{\' o}pez (2004) examine the effects of the fast wind on
a clumpy AGB wind which is more realistic than the smooth density
distributions usually considered in the IWs models.

There is now an abundance of observational 
evidence that  fast winds exist within PNe and some evidence that they
interact significantly with the circumstellar envelopes. 
Patriarchi \& Perinotto (1991) discovered that 60 percent of central stars
of PNe emit particle winds of 600--3500 \kms. However, direct observational
evidence of their interaction with the circumstellar medium is more
limited. Collimated and truncated  
ablated flows, where the fast wind has mixed
with, and is slowed by, photoionised gas evaporating from dense, stationary, 
globules (Hartquist et al. 1986; Dyson et al. 1989a; Dyson, Hartquist
\& Biro 1993) are detected
in the hydrogen--deficient PNe A30 (Borkowski et al. 1995;
 Meaburn \& L{\'o}pez,1996) and A78
(Meaburn et al. 1998). Also, diffuse X--ray emission is found in the cores
of five PNe ( NGC 7009, Hen 3--1475, BD+30$\deg$ 3639,
NGC 6543 \& NGC 2392 -- Chu et al. 2001;
Gruendl et al. 2001
Guerrero, Chu \& Gruendl 2004; Chu et al. 2004; 
Guerrero
et al. 2005). Similarly, Kastner
et al. (2003) observed such diffuse X--ray emission in the core of the
bi--polar PN Menzel 3 (Mz~3)
and Montez et al. (2005) inside the main shell of NGC 40
which is a PN generated by a WR--type star. 
All of these authors interpret
the X--ray emissions to be the consequences of the collisions of the 
fast winds with the slower moving surrounding AGB winds. 
They suggest that conductive cooling is occurring for the temperatures 
of the hot gases
emitting the X-rays are far lower than if simply generated by shocks
in the fast winds at their measured speeds. Nonetheless, they imply
that over--pressured `bubbles' of super-heated gases are forming 
and driving the expansions of the ionised PNe shells as predicted by
the IWs model.

The principal purpose of the present article is to examine,
on the basis of observations
made recently with the two Manchester echelle spectrometers (MES - Meaburn
et al. 1984 and 2003), the part played 
by the fast winds in
the creation of the well--evolved PNe, NGC 6853 (Dumbbell) and 7293 (Helix),
the young PN, NGC~6543 
and the
outer lobes of the
bi--polar (poly--polar) `PNe' NGC 6302, Mz~3 and MyCn~18 for 
these are all recently observed examples of the range of circumstellar
phenomena broadly designated as PNe.

\section{The evolved PNe, NGC 6853 and 7293}

Cerruti--Sola \& Perinotto (1985) and Patriarchi \& Perinotto (1991)
failed to detect any fast winds of $\geq$ 10$^{-10}$ \msun\ yr$^{-1}$
from either of the central stars of NGC 6853 and 7293 in their IUE
observations. This is not surprising for both stars are well into their
WD phases with surface temperatures $\approx$ 10$^{5}$ K ( G{\'o}rny, 
Stasi{\'n}ska \& Tylenda 1997; Napiwotzki 1999) and well past the
transitions from their AGB to WD phases during which periods 
the fast winds are expected to blow.
The question is, do the present morphologies and kinematics of these
PNe  depend critically on the previous emissions of  fast winds
if they in fact occurred? 
Observations of NGC 6853 are reported in Meaburn et al. (2005a) 
and of NGC 7293 in Meaburn et al
(2005b) in an attempt to throw light on this question. These
should be combined with an appreciation of
the complementary imagery for NGC 7293 of O'Dell (1998) and
O'Dell, McCullough \& Meixner (2004). 

\begin{figure}
\centering
\includegraphics[height=12cm]{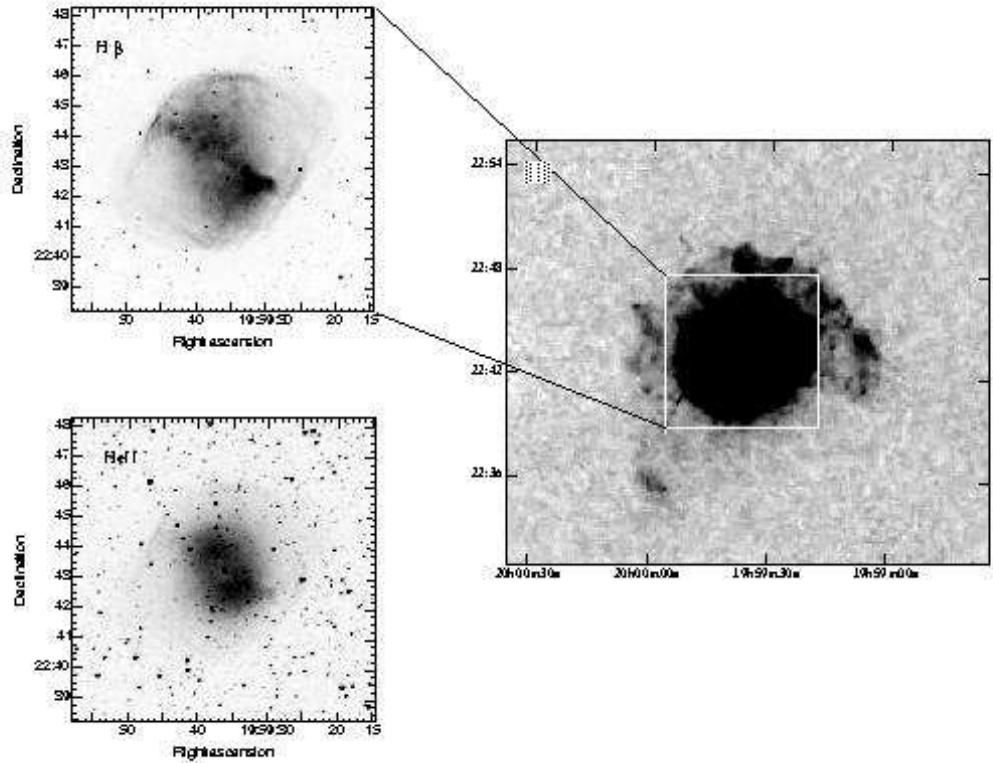}
%
%
\caption{ A deep \ha\ + \nii\ image of NGC 6853 taken by Panos Boumis
is compared with
\hb\ and \heiis\ images (compliments of Bob O'Dell) of the familiar
`Dumbbell' nebula (coords throughout are for J2000)
}
      
\end{figure}

\begin{figure}
\centering
\includegraphics[height=10cm]{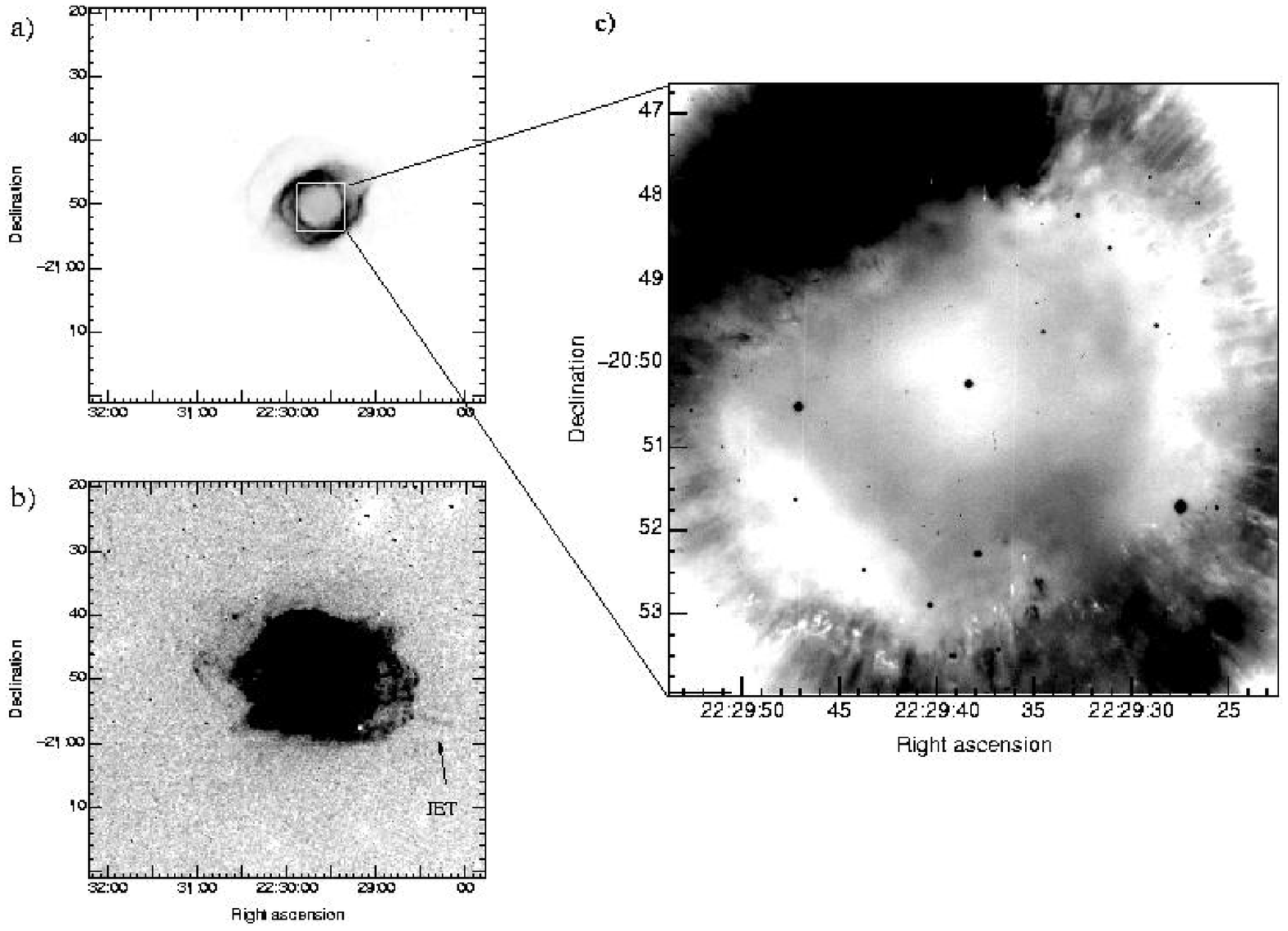}
\caption{The \ha\ + \nii\ image in a) shows the familiar, bright
helical structure which gives the nebula its name `Helix'. It is shown
in c) that this is enveloped in complex outer halos that may even include
a jet. The NTT (La Silla) \oiiil\ image in c) reveals 
particularly well the inner 
\oiiil\ shell.  }
\end{figure}

Some aspects of the ionisation stratification of NGC 6853 and 7293 
can be appreciated in Figs. 1, 2 \& 3 and of the corresponding velocity
structure in Figs.3 and  4a \& b. Highly excited gas emitting 
\heii\ lines is expanding slowly
around both exciting stars. These  central volumes are themselves
both  surrounded by 
faster shells of lower excitation emitting the  \oiii\ lines.
All are contained within outer and even faster expanding lowly 
ionized shells emitting the \nii\ lines.For NGC 6853 the central \heii\ volume
(0.38 x 0.33 pc$^{2}$) is expanding at $\leq$ 7 \kms, the inner
\oiii\ shell at 13 \kms\ and the outer ellipsoidal (0.50 x 0.67 pc$^{2}$)
\nii\ shell at 35 \kms. For NGC 7293 the central \heii\ volume (0.21 pc diam.)
is expanding at $\leq$ 11 \kms\ , the inner \oiii\ shell (0.25 pc. diam.)
at 12 \kms\ and the outer \nii\ structure (0.64 pc. across and 
shown in Meaburn et al. 2005b to
be  bipolar with an  axis  tilted at 37\degree\ to the sight line 
to give the characteristic
helical appearance of NGC 7293) is expanding at 25 \kms.
The deep images in Figs. 1 and 2 show the bright regions of both
nebulae are surrounded by clumpy haloes which could be, within the IWs model,
the as yet unaccelerated AGB wind but could alternatively be the prior RG wind.

\begin{figure}
\centering
\includegraphics[height=10cm]{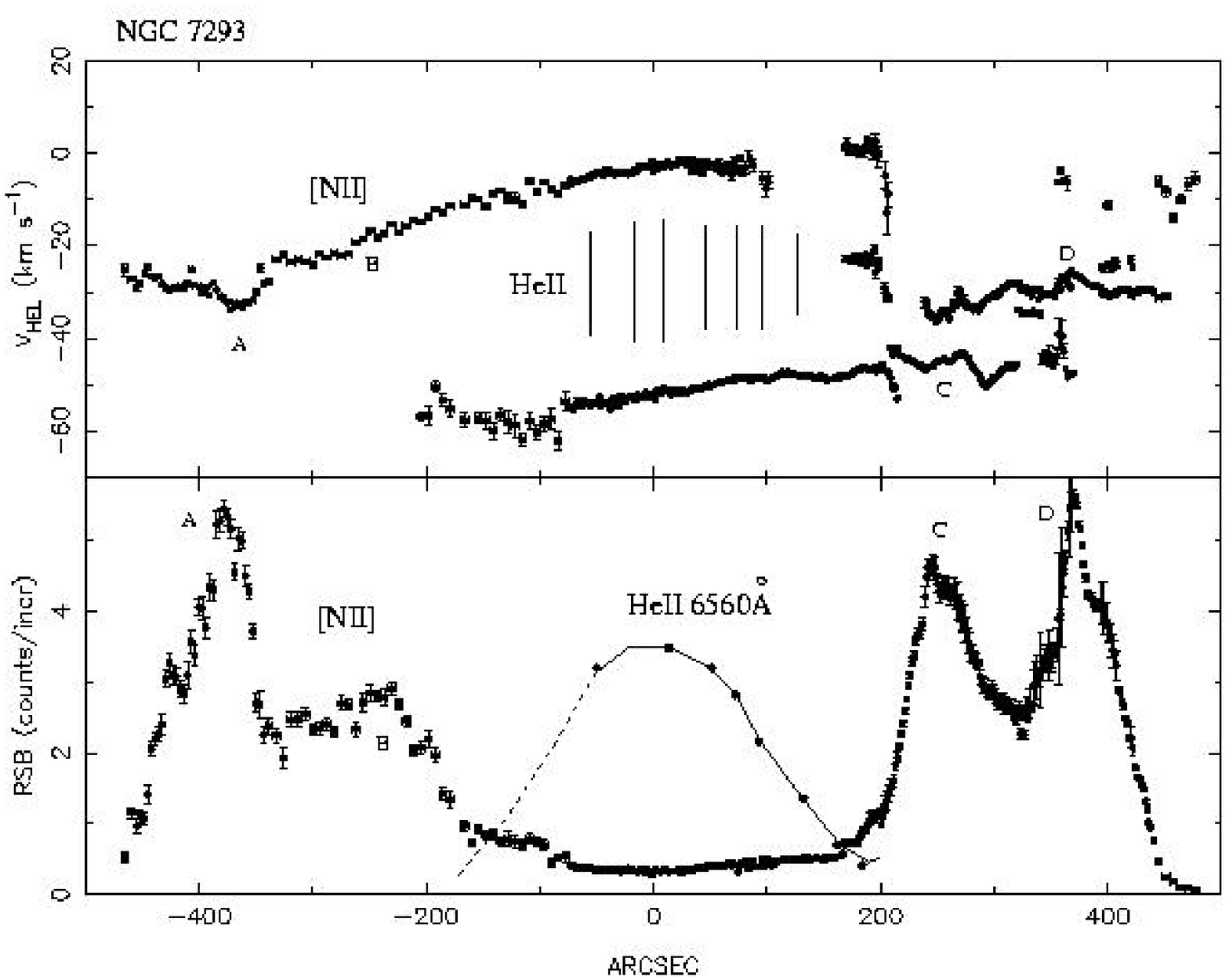}
\caption{In the top panel the radial velocities of separate velocity
components in the \niil\ profiles along an 
EW cut through the central star and over the helical structure in Fig. 2a 
are compared
with the widths of single Gaussians that simulate  the \heiil\ profiles 
around the nebular core. The latter have been corrected for instrumental
broadenings. The relative surface brightness (RSB) variations of the
\niil\ and \heiil\ profiles are shown
in the bottom panel along the same cut. The brightness peaks A-D over the
helical structure are marked.  }
\end{figure}

\begin{figure}
\centering
\includegraphics[height=12cm]{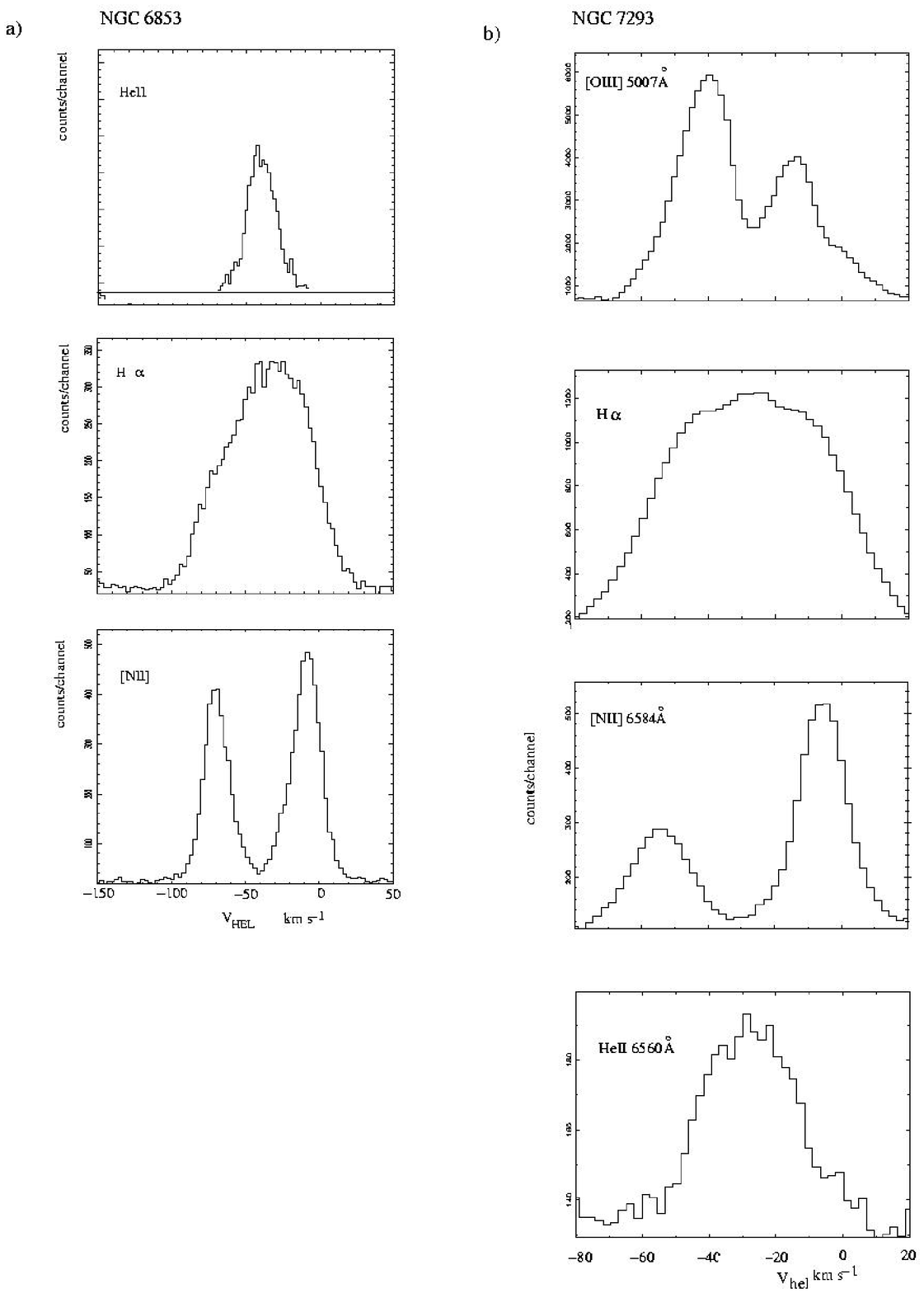}
\caption{a) Line profiles over the core of NGC 6853 are compared. c) As for
a) but for NGC 7293.}
\end{figure}

Even if there were fast winds present in NGC 6853 and 7293 they could not
reach the outer shells unless they percolated as mass--loaded flows through
clumpy inner \heii\ and \oiii\ emitting regions (Meaburn \& White, 1982). 
However, 
with no current fast wind observed
the IWs model could then only apply in the earliest post--AGB stages 
of the formation of these PNe i.e. the fast winds initially formed  expanding
bubbles then switched off.
A somewhat elaborate consequence within the IWs model for an evolved PN 
could be the acceleration inwards
of the inside surface of the outer shell, for it would no longer
be supported by the over-pressure of the super--heated shocked gas, 
to cause the slower moving
inner regions reported in Meaburn et al. (2005 a \& b). The gap between
the outer edge of the \oiii\ inner shell and the inside edge of the
\oiii\ and \nii\ outer shell (Fig. 2) is not simply explained by
this inward acceleration model.

Alternatively, the fast wind never existed or was
ineffective over the long term. The expanding shells could then simply be 
pulsed,
higher speed, ejections of AGB wind (for not understood reasons). 
The dynamical
ages  of the central \heii\ volume, the inner \oiii\ and the 
outer \oiii\ plus \nii\ NGC 7293 shells are all
$\approx$ 10$^{4}$ yr and the central \heii\ volume and  
outer \nii\ shell of NGC 6853 
 $\approx$ 8000 yr.
Within this situation, in each PN, all of these emitting regions 
would have been ejected at about 
the same time but with decreasing velocities i.e. Hubble--type outflows.
The haloes of NGC 6853 and 7293 could still be the prior, but lower speed,
AGB or even RG winds emitted over $\geq$ 10$^{5}$ yr.

A comment must be made about the tails of the cometary knots in NGC 7293. As 
the fast wind is currently not observed in this PN 
it cannot be the cause of these
radial tails. An alternative remains that dense knots in the AGB (or even 
RG wind)
seen initially as SiO maser spots (Dyson et al. 1989b) 
are overrun by a pulse of AGB 
wind to draw these tails out. 

\section{The young PN, NGC 6543}

The `Cat's Eye' nebula, NGC 6543,
is a young PN photoionised by an O7+WR--type star which emits
a high--speed particle wind at 1900 \kms (Patriarchi \& Perinotto, 1991). 
Its bright filamentary core is complex but within an overall 
25 arcsec $\times$ 17 arcsec ellipse ($\equiv$ 0.12 pc $\times$ 0.08 pc
for a distance of 1001 $\pm$ 269 pc as given by Reed et al. 1999). This
core is surrounded by a highly filamentary structure, 330 arcsec diam.
($\equiv$ 1.6 pc and see the image by Romani Corradi in Mitchell et al.
2005). Chu et al. (2001) have shown conclusively that diffuse X--ray
emission is confined within an `inner' ellipse of optical line emission,
with a minor axis of 8 arcsec across, itself embedded in the larger
bright core. Miranda and Solf (1992) had shown that this inner elliptical
feature is expanding at 16 \kms\ to give a dynamical age of 2400 yr.
The fast wind must therefore be confined to this small inner region.
The larger filamentary features in the core, which surround this inner
ellipse, and the bi--polar jets, seem independent of the presence of the
fast wind and are most likely ejecta.

Bryce et al. (1992) had demonstrated that the outer, high excitation,
halo is very inert
i.e. expanding globally at 4.5 \kms\ which gives it a dynamical age
of 1.7 $\times$ 10$^{5}$ yr. Mitchell et al. (2005) show that all of
the flows off globules in this halo are around the sound speed
and therefore are soley a consequence of ionisation fronts created
by photoionisation. There is no evidence of any interaction
with the fast wind which must be confined to the inner ellipsoidal
shell within the nebular core.

It is concluded that the post--AGB phase started only around 2400 yr
ago and that a small shell, predicted by the IWs model is being
driven by the fast wind into an extremely clumpy AGB wind, with
maybe the outer halo even being the slow moving relic of the most
recent RG wind. If the fast wind were to blow for 10 times its present
age it is difficult to visualise the creation of a large expanding
shell in such a clumpy outer halo.

%
%
%

\section{The bi--polar PNe, NGC 6302, Mz 3 and MyCn 18}

These bi--polar nebulae must have more complicated 
stellar systems than NGC~6853, 7293 and 6543 and as well
have circumstellar disks. Here, following Bains et al. (2004), Smith (2003)
and Smith \& Gehrz (2005), they are
designated PNe for they have, arguably, post--AGB elements in their nature
although  the central stars are most likely close binaries.

\subsection{NGC 6302 -- a high excitation poly--polar PN} 

NGC~6302 (PN~G349.5+01.0) is a poly-polar planetary nebula (PN), which
was described and drawn as early as 1907 by
Barnard. It is in the highest excitation class of PNe with a
central
O {\sc vi}--Type White Dwarf and possible binary companion (Feibelman 2001).
This stellar system is heavily obscured by a dense circumstellar disk 
(Matsuura et al. 2005).

\begin{figure}
\centering
\includegraphics[height=12cm]{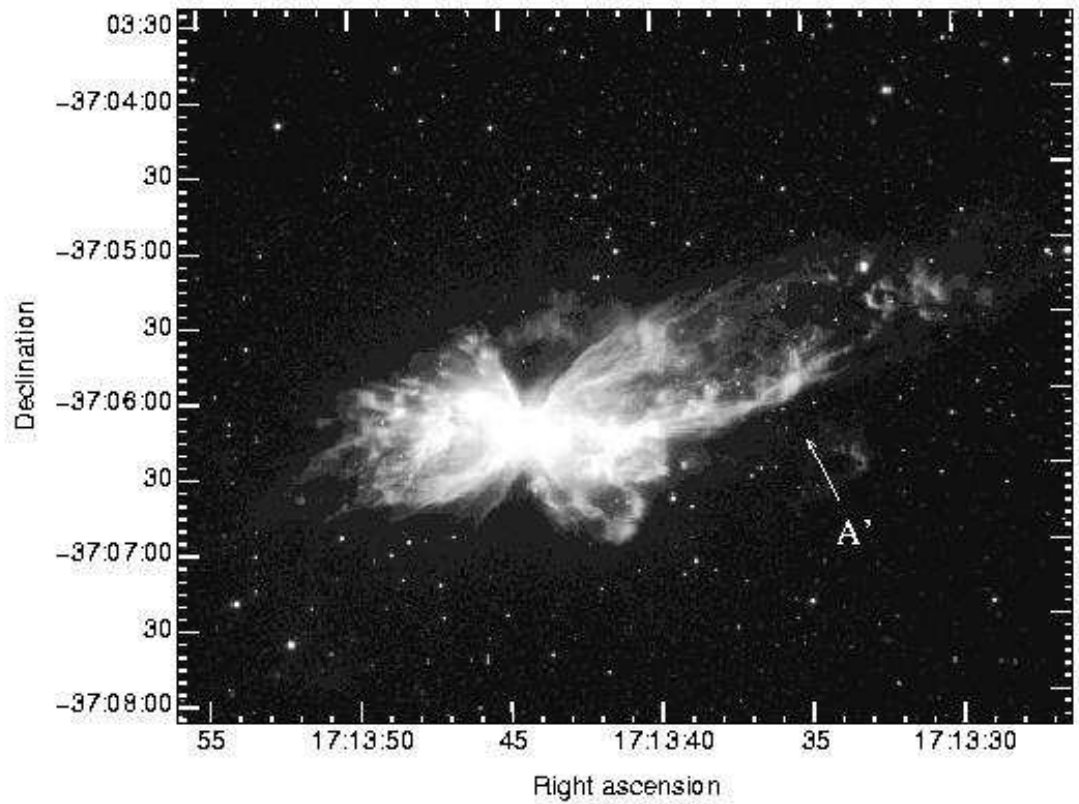}
\caption{An \ha\ + \nii\ image of NGC 6302 taken by Romani Corradi
with the 3.6--m La Silla telescope. The cut A' shown in Fig. 7 is marked.}
\end{figure}

\begin{figure}
\centering
\includegraphics[height=14cm]{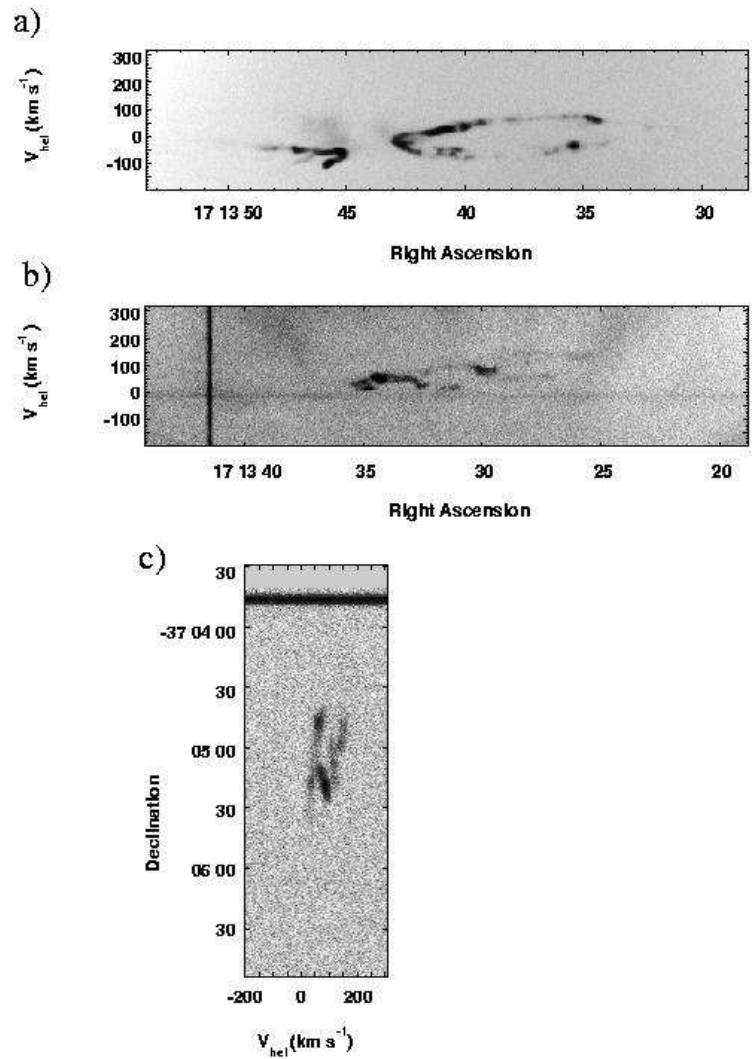}
\caption{Sample PV arrays of \niil\ line profiles over NGC 6302 are
shown. Those in a) and b) are for an EW slit centred on DECs 
-37 05 50 and -37 05 02  respectively.
That in c) is a NS slit centred on RA 17 13 28. By 
comparison with the image in
Fig. 5 it can be seen that the arrays in b) and c) cover the extremities
of the prominent NW lobe of NGC 6302.}
\end{figure}

\begin{figure}
\centering
\includegraphics[height=14cm]{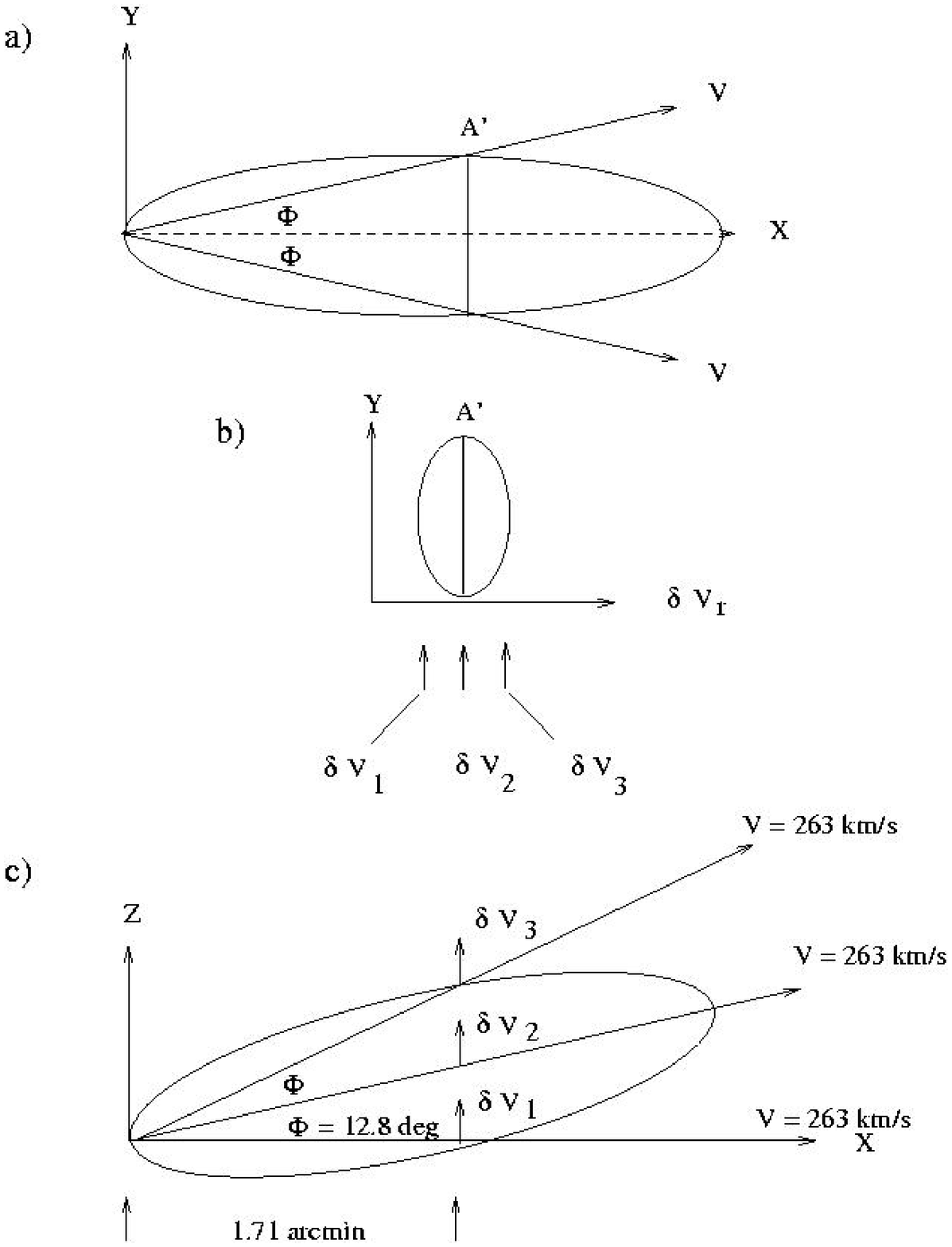}
\caption{The image of the NW lobe of NGC 6302 is shown schematically 
as an ellipsoidal 
structure with circular section
in a) where X and Y are in the plane of the sky. The expansion velocity V
is shown to be Hubble--type. The velocity ellipse found along cut A' (1.71
arcmin from the central star) in
Fig. 5 is from Meaburn \& Walsh (1980) and gives the parameters shown
in c) where the Z dimension 
is perpendicular to the plane of the sky (the observer is below c). 
With Hubble--type
expansion, V $\approx$ 600 \kms\ at the extremity of the NW lobe
(see Fig. 6b).}
\end{figure}

The kinematics of the prominent NW lobe of NGC~6302 (Fig. 5) have been
determined in detail by Meaburn \& Walsh (1980) and most recently by
Meaburn et al. (2005c) -- see examples from the latter in Fig. 6. Meaburn \&
Walsh (1980) showed `velocity ellipses' in the position--velocity (PV)
arrays of line profiles across
the diameter of the lobe (as sketched in Fig. 7b and see Meaburn et al.~2005c)
showed that this outflow is Hubble--type reaching V = 600 \kms\ at the 
extremities of the lobe (Figs. 6b \& c). A `spot' value of V = 263 \kms\
at position A' in Fig. 5 (1.71 arcmin from the star) is shown in Fig. 7c.
No fast wind has been directly observed (Feibelman 2001) from the 
O VI--Type WD star and its possible 
companion. For an expansion--proper motion distance of 
1.04 $\pm$ 0.16 kpc (Meaburn et al. 2005c)
the dynamical age of the lobe is 1900 yr.

It is concluded in Meaburn et al. (2005c) that an eruptive event 1,900 yr ago 
created the prominent NW lobe
and possibly many of the other lobes.

\subsection{Mz 3 -- a symbiotic PN}

Bains et al. (2004) and Smith (2003) suggest that the central
stellar system of the PN, Mz~3, (Fig. 8) is a symbiotic binary. 
L{\'o}pez \& Meaburn (1983) had shown that 
the bright central bi--polar shells
of Mz 3 (the 9 arcsec diam. N shell and 
14 arcsec diam. S shell in Fig. 8) are in spherical expansion at 
40 \kms\ and 55 \kms\ respectively. 
For a distance to Mz 3 of 1.3 kpc (see Bains et al. 2004 for a 
review of possible distances) a mean dynamical age of 1435 yr for
these inner shells is implied. These shells are on either
side of a dense disk (Meaburn \& Walsh 1985) which obscures the
central stellar system. 

\begin{figure}
\centering
\includegraphics[height=12cm]{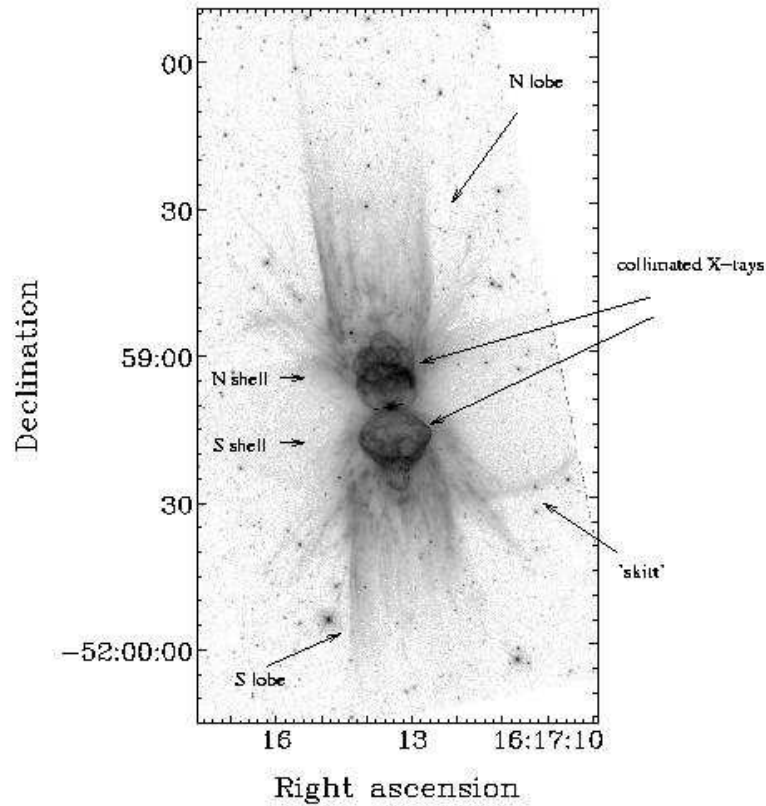}
\caption{An HST image of Mz--3 in the light of \ha\ + \nii.
The N and S lobes exhibit Hubble--type outflows and the central
bright shells, N and S shell,
contain diffuse X--ray emission though collimated along
the lobe axis. The high speed `skirt' identified by Santander--Garcia et al.
(2004) is marked.}
\end{figure}

Furthermore, Meaburn \& Walsh (1985) revealed that the N and S lobes in Fig. 8
had circular sections, for velocity ellipses occurred in the PV arrays
of line profiles over their diameters, and that their outflows 
at velocity V along
vectors  directed away from the central star (as for NGC~6302 in Fig. 7
though with different angles) 
are 
very similar to those of NGC 6302 and similarly
Hubble--type. Spot values of V at 20 arcsec N and S of the central
star are given by Meaburn \& Walsh (1985) as 90 and 93 \kms\
respectively. As for NGC 6302 (Sect.3.1) the Hubble--type nature
of the outflows implies that at the limits of detection of these
lobes (82 arcsec N and 52 arcsec S) V  reaches 369 \kms\ and 
242 \kms\
respectively as confirmed by Santander-Gracia et al. (2004). The corresponding
dynamical ages for the N and S lobes of Mz 3 are consequently
1317 and 1273 yr respectively which are remarkably similar to
those of the N and S shells.

The implication is that {\it all} of the outflows of the N and S shells and
the N and S lobes are within a general Hubble--type velocity system
which reinforces the possibility 
that they are all the consequence of ejections at
closely similar times but with different ejection
speeds. In any case the fast wind must be
contained within the N and S shells to shield the N and S lobes from
any direct interaction.
Furthermore, the X--ray emission (Kastner et al. 2003) suggests 
that this fast wind is 
collimated along the bi--polar axis of Mz 3 and is causing the secondary
protusions at the apices of the N and S shells evident in Fig. 8.
This would preclude the N and S shells themselves from being 
`bubbles' driven only by this fast wind.

Incidentally, the high--speed skirt (Fig. 8) has been properly
identified 
by Santander-Garcia et al. (2004)
as the origin of the high--speed `velocity ellipse'
in the PV arrays of Meaburn and Walsh (1985) and the high--speed
feature in Redman et al. (2000).

\subsection{MyCn 18 -- a nova--like PN} 

 MyCn~18 is also aptly known as the `Engraved Hourglass' nebula due to
the visually dramatic bipolar appearance of its bright core
(Sahai~et~al.1999 and references
therein). 
However, interest in MyCn~18 has
recently been further heightened by the discovery of the knots of ionized
gas flowing in both directions along its bipolar axis at speeds of up to
660 \kms
(Bryce et al. 1997: O'Connor et al. 2000). These can be seen in the continuum
subtracted image in Fig.9a.
Corradi~\&~Schwarz~(1993) had previously
investigated the bright core of MyCn18 and concluded that it is a
young PN. The presence of a dusty, molecular, equatorial waist region,
suspected by Sahai~et~al.~(1999) on the basis of an excess in the
stellar K-band photospheric flux, substantiated this young age.  The
radio thermal emission map of
Bains~\&~Bryce~(1997) reveals the ionised inside
surface of the dense waist region to be very bright in comparison to
emission from polar directions.

\begin{figure}
\centering
\includegraphics[height=16cm]{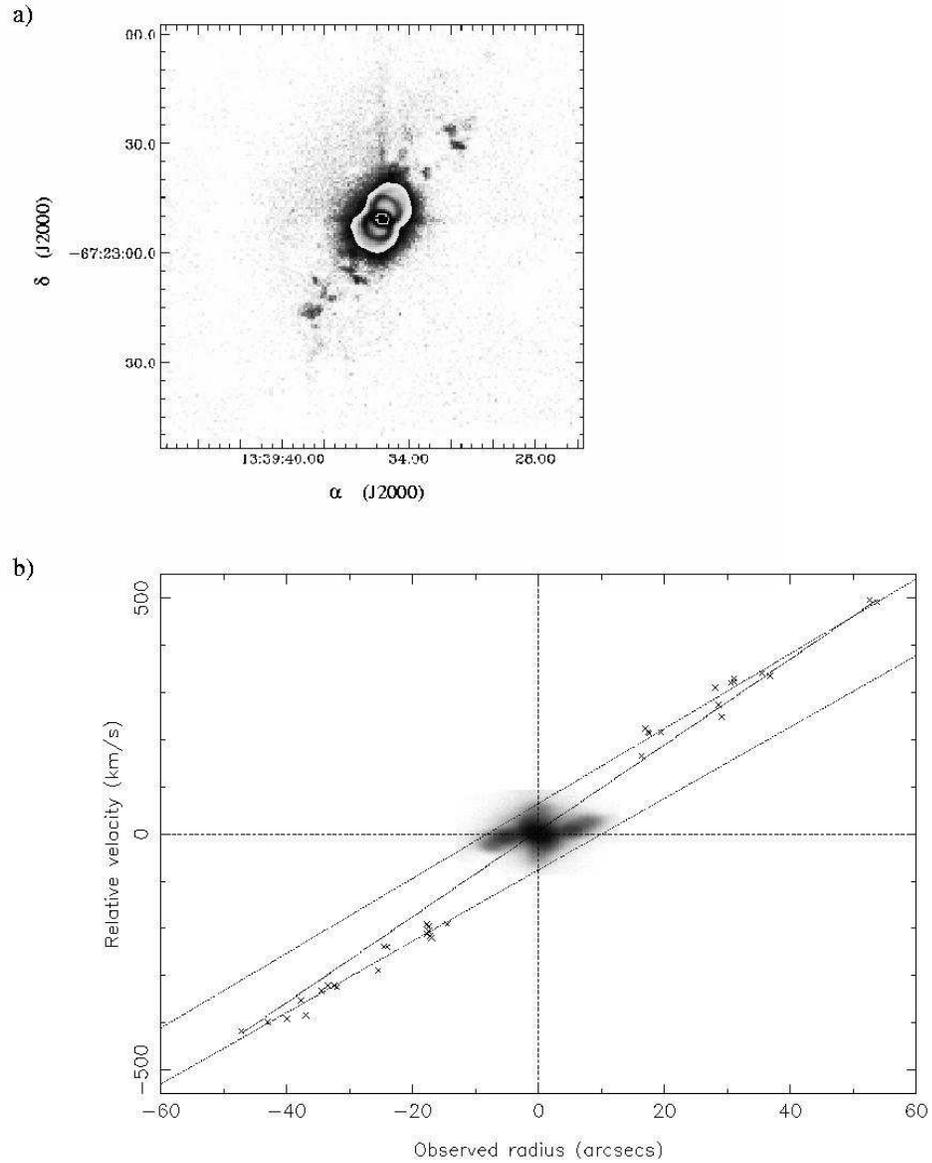}
\caption{In a) the image of the bright core of MyCn 18 taken with
the AAT (Bryce et al. 1997 and O'Connor et al. 2000) is shown
compared with the faint knots along the lobe axis. In the PV diagram in 
b) the relative radial velocities of these knots
are shown as crosses. Very high--speed, Hubble--type,  bi--polar motions
are indicated for the radial velocities follow straight lines.}
\end{figure}

The Hubble-type nature of the knotty outflow is clear in 
Fig. 9b (from O'Connor et al. 2000). It is notable that the best fit
straight lines are significantly displaced from the systemic
radial velocity near the nebular core. O'Connor et al. (2000) show
that these knots were ejected over a 300 yr period with a dynamical
age of 1250 yr (Bryce et al. 1997). They also conclude that
dynamically, the most plausible explanation
seems to be that the high speed knotty outflow from MyCn18 is the
result of a (possibly recurrent) nova--like ejection from a central binary
system. This is in harmony with the considerations of
Sahai~et~al. (1999) who favour a close binary system to generate the
morphology of the very innermost regions of MyCn~18.
In these circumstances the knots would be the manifestation of
ballistic, dense, bullets ejected with a range of speeds
and not the consequence of acceleration by a fast wind.

\section{Conclusions}
The IWs model certainly seems applicable to the very innermost shell of the 
young PN NGC 6543 (and very clearly to the main 
shell of NGC 40 -- see Sect.~1).
However, considerable modification of this theory is needed to explain the
current state of the 
evolved PNe, NGC 6853 and 7293. The fast wind could have switched off
well before the 10$^{4}$ yr age of their expanding shells 
generating an inward acceleration of their inside surfaces. 
Alternatively, it remains possible that eruptive events over 10$^{4}$ yr have
dominated any effects of the fast winds, active only for a few
thousand years in these PNe, to create the Hubble--type expansions
throughout their volumes that are currently observed.

The preponderance of Hubble--type outflows of the lobes of the
bi--polar PNe, NGC 6302, Mz 3 and MyCn 18 (and see Corradi 2004 for
other examples) invites the simplest interpretation; that they are
consequences of ejections over short periods of time but of
material with different speeds. Even when a fast wind
is currently present (e.g. Mz 3) the outer high--speed 
lobes are shielded from it
by inner shells.

The present article has been deliberately limited to considering the dynamics
of a small number of objects whose motions have been well observed. It seems
clear in this small sample that simple ballistic ejections, 
maybe involving close
binary systems in some cases, could dominate the dynamical effects
of the fast winds. 

The author is grateful to Bob O'Dell for 
providing the \hb\ and \heiis\ images in Fig. 1 and to Romani
Corradi for the image in Fig. 5.

%


\printindex
\end{document}